\documentclass[prd, amsfonts, twocolumn, nofootinbib, showpacs]{revtex4}
\usepackage{graphicx, epsfig}
\usepackage{color}
\usepackage{amsmath, cancel}
\newcommand{\be}{\begin{equation}}
\newcommand{\ee}{\end{equation}}
\newcommand{\bea}{\begin{eqnarray}}
\newcommand{\eea}{\end{eqnarray}}

\newcommand{\gapp}{\mathrel{\raise.3ex\hbox{$>$}\mkern-14mu
\lower0.6ex\hbox{$\sim$}}}
\newcommand{\lapp}{\mathrel{\raise.3ex\hbox{$<$}\mkern-14mu
\lower0.6ex\hbox{$\sim$}}}
\def\bbox{{\,\lower0.9pt\vbox{\hrule \hbox{\vrule height 0.2 cm
\hskip 0.2 cm \vrule  height 0.2 cm}\hrule}\,}}

\begin{document}
\title{Icezones instead of firewalls: extended entanglement beyond the event horizon and unitary evaporation of a black hole}
\author{John Hutchinson$^{1,2}$\footnote{Previous name John Wang}, Dejan Stojkovic$^1$}
\affiliation{ $^1$ HEPCOS, Department of Physics, SUNY at Buffalo, Buffalo, NY 14260-1500}
\affiliation{ $^2$ Department of Physics, Niagara University, NY 14109-2044}


\begin{abstract}
\widetext
We examine the basic assumptions in the original setup of the firewall paradox. The main claim is that a single mode of the lathe radiation is maximally entangled with the mode inside the horizon and simultaneously with the modes of early Hawking radiation. We argue that this situation never happens during the evolution of a black hole. Quantum mechanics tells us that while the black hole exists, unitary evolution maximally entangles a late mode located just outside the horizon with a combination of early radiation and black hole states, instead of either of them separately. One of the reasons for this is that the black hole radiation is not random and strongly depends on the geometry and charge of the black hole, as detailed numerical calculations of Hawking evaporation clearly show. As a consequence, one can't factor out the state of the black hole. However, this extended entanglement between the black hole and modes of early and late radiation indicates that, as the black hole ages, the local Rindler horizon is modified out to macroscopic distances from the black hole.  Fundamentally non-local physics nor firewalls are not necessary to explain this result.   We propose an infrared mechanism called {\it icezone} that is mediated by low energy interacting modes and acts near any event horizon to entangle states separated by long distances.  These interactions at first provide small corrections to the thermal Hawking radiation.  At the end of evaporation however the effect of interactions is as large as the Hawking radiation and information is recovered for an outside observer.  We verify this in an explicit construction and calculation of the density matrix of a spin model.
\end{abstract}


\maketitle

\section{Introduction}

The authors of Ref.~\cite{Almheiri:2012rt} challenged our understanding of black holes by claiming that for old black holes, quantum mechanics and the usual black hole postulates are not compatible.  Keeping quantum mechanics unchanged, they examined the following statements about black holes:  (1) Hawking evaporation is information preserving, (2) low energy effective field theory should be valid beyond some microscopic distance from the horizon, and (3) an infalling observer does not see anything unusual at the horizon.  In particular, they examined a mode of late Hawking radiation, which we call $L$, emitted by an old black hole.  They argued that this mode is maximally entangled with earlier radiation, which we call $E$, due to general arguments by Page about thermodynamic systems.   Moreover they claimed that the mode $L$  should be entangled with a mode, $L^\prime$ inside the horizon because the horizon is smooth for any infalling observer.
Having a mode $L$ maximally entangled with two different states would represent a paradox because maximal entanglement is monogamous, namely one state can not be simultaneously maximally entangled with two different states.
 To resolve the paradox, the authors proposed their most conservative resolution claiming that black hole statement $3$ is incorrect so an infalling observer near the horizon encounters some new and unknown high energy physics dubbed a firewall that burns infalling objects.  According to their arguments any alternative would be much more radical and would involve violations of semiclassical physics out to macroscopic distances from the horizon.

In this paper we first re-examine the entanglements in a system using Page's result \cite{Page:1993df} which applies generically to thermodynamic systems.  If a system evolves into a large subsystem of dimension $n$, and a small subsystem of dimension $m$, then eventually there is almost no information left in the small subsystem.
Page defines information as the difference between the maximal and average entropy in the system, i.e. $I= S_{\rm max} - <S>$, and shows that information contained in the small system is about $m/(2n)$, which means that the small system contains less than half a bit of information. Instead, all of the information is contained in subtle correlations between the small and large sub-systems.

If the large subsystem is early Hawking radiation and the small subsystem is late Hawking radiation, this result is often interpreted as the fact that early radiation and a mode of the late radiation must be maximally entangled.

As we will discuss in more detail in Sections  \ref{u} and \ref{I}, a mode of late radiation is maximally entangled with the rest of the large subsystem, which after the black hole has evaporated, contains both the early radiation and the rest of the late radiation.  Page's result can additionally be applied at different times during the evaporation process.  For example when we have an old black hole, we can take the small subsystem to consist of a single mode of the late Hawking radiation. Because of unitarity, this mode should be entangled with both the early radiation and some black hole states (despite the fact that we usually trace over the states which are inside the black hole).  Because the exterior mode and the interior mode are not maximally entangled,
then there is no need for a firewall to destroy entanglements.  Our resolution implies that the horizon modes of an old black hole are strongly entangled with the early Hawking radiation, while the horizon modes are more strongly entangled with each other for a young black hole.  As the black hole ages, the horizon modes become increasingly entangled with the early Hawking radiation due to an infrared effect that we call {\it{icezone}}.  While the infrared dynamics are local, over a period of time their accumulated effect increasingly entangles states out to macroscopic distances from the horizon.  Quantum unitarity is preserved.

In Section \ref{dm} we quantify the icezone by constructing its density matrix.  As an explicit example of how icezone works in practice we study a particular spin emission model of black hole evaporation where interactions change an approximately thermal density matrix into a pure state.  These interactions introduce small corrections to Hawking radiation during most of the decay process.  They become increasingly important at the end of evaporation and unitarize the radiation.

\section{Entanglement in Unitary Model of Black Hole Decay}
\label{u}

In this section and section \ref{I} we examine the original assumptions of the paradox itself.  Specifically, we point out that unitarity of the black hole evaporation process implies that, $L$, a mode of late Hawking radiation, is maximally entangled with the combination of early Hawking radiation $E$ and black hole states while the black hole exists, and not with either of them separately.  The late mode $L$ is not maximally entangled with an interior mode at any time except maybe for very young black holes.   At the end of evaporation, $L$ is maximally entangled with the early Hawking radiation and even later Hawking radiation.  The late Hawking mode is not maximally entangled with just the early Hawking radiation unless $L$ is the last mode emitted. Thus, the basic premise of the firewall paradox, i.e. a double maximal entanglement, is incorrect.

\begin{figure}[h]
  \centering
\includegraphics[width=9cm]{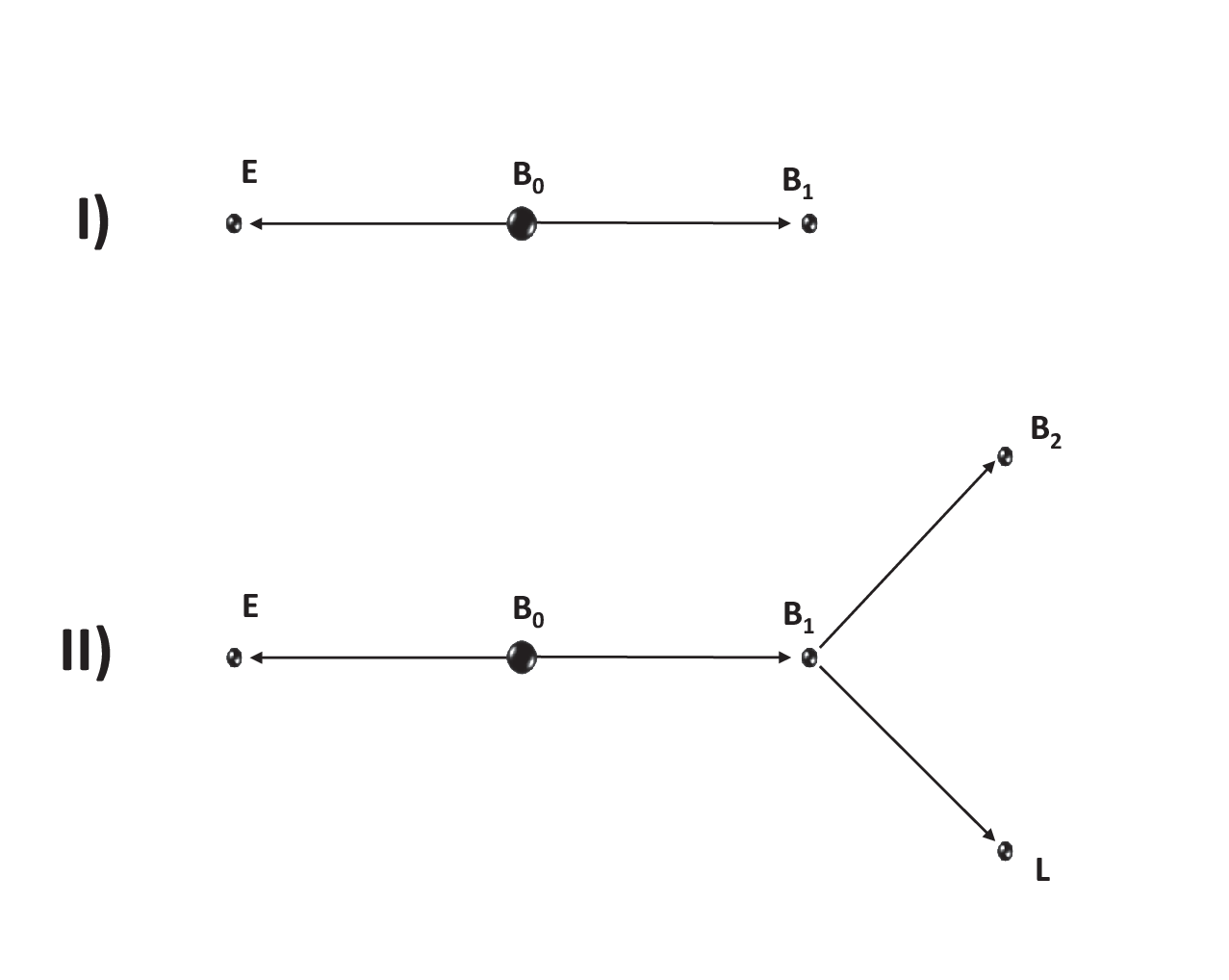}
\caption{This picture illustrates the unitary decay on an initial state $B_0$. The decay products $B_1$ and $E$ are typically in a highly entangled state because of their common origin.  The state $E$ is highly entangled with the pair ($B_2+L$), but there is no reason to expect that $E$ is highly entangled with $B_2$ and $L$ separately. If $B_0$ is an initial black hole that emits early radiation $E$ and changes its state into $B_1$, and then $B_1$ emits the late radiation $L$ and changes its state to $B_2$, we have a unitary model of black hole evaporation. Therefore, $E$ and $L$ are never maximally entangled, $E\cancel{\leftrightarrow} L$, except at the end of the process when there is no black hole.
Thus, the basic premise of the firewall paradox, i.e. a double maximal entanglement, is incorrect.}
    \label{process}
\end{figure}

Black hole evaporation resides in the still swampy lands of quantum gravity.  To understand black hole decay we start from terra firma and try to navigate as far as possible using the rules of quantum mechanics.  Let us begin with a general two step quantum mechanical decay process that consists of a pure state $B_0$ decaying into two states, $B_1$ and E as shown in Fig.~\ref{process}.  The decay products $B_1$ and $E$ are typically in a highly entangled state because of their common origin.  Next, the state $B_1$ decays into two other states, $B_2$ and $L$. Since the state $E$ was entangled with $B_1$, and $B_1$ yielded $B_2$ and $L$, we also expect that the state E is highly entangled with the pair ($B_2+L$).  There is no reason to expect that $E$ is highly entangled with $B_2$ and $L$ separately, and such a situation would violate entanglement monogamy.  On the other hand a high level of entanglement between $E$ and the pair ($B_2+L$) is expected and does not violate any principles.

To map this decay process to the firewall paradox, take $B_0$ to be a black hole that emits early Hawking radiation $E$, while $B_1$ represents a less massive black hole that emits late Hawking radiation mode $L$.
Therefore, when a black hole $B_0$ emits an early Hawking mode $E$, the resulting state consists of $B_1$ and $E$ which are maximally entangled $B_1\leftrightarrow E$.  When subsequently $B_1$ emits a late Hawking mode $L$, the resulting state consists of $B_2$ and $L$.  In this case the late radiation $L$ is not maximally entangled with the early radiation $E$ or with any modes inside the black hole separately.  The late mode $L$ is only maximally entangled with a combination of the black hole and early radiation $E\leftrightarrow (B_2+L)$.  The decay process can be summarized in the following steps
\be
\begin{bmatrix}
step&state&entangled\ states  \\
0&B_0&  \\
1&B_1+E & E \leftrightarrow B_1\\
2&B_2+E+L& E \leftrightarrow (B_2+L)
\end{bmatrix} \ .
\ee
By construction the process described in Fig.~\ref{process} does not represent a paradox, so perhaps the firewall setup does not represent a paradox either.

Let's connect our general discussion to a specific example. To discuss entanglement we have to label our states somehow. For simplicity, we use spin. We start with a black hole whose initial spin is zero. We label this state as $B_0$.  In step one, the black hole emits an electron. This electron represents early Hawking radiation. The system consisting of a black hole plus the emitted electron is now in the entangled state described as
\be \label{be}
\psi_{B_1 E} = \alpha \left| \uparrow_{B_1} \ \downarrow_E \right\rangle  + \beta \left| \downarrow_{B_1} \ \uparrow_E \right\rangle
\ee
where the subscripts B and E stand for ``black hole'' and ``early Hawking radiation''.  Black holes conserve angular  momentum (for simplicity spin) during Hawking radiation, so if a spin down electron is emitted, the black hole has to acquire one unit of spin up, and vice versa.  Before we do the measurement and collapse the wave function of the emitted electron, the system ``black hole + early Hawking radiation" is in the entangled state (\ref{be}).

In step two, the black hole emits another electron, which now represents late Hawking radiation.  The electron emitted later is entangled with the ``black hole + early Hawking radiation'' state, and it would wrong to consider the entanglement of the black hole with late radiation, and late radiation with early radiation separately. The correct entangled state of the system ``black hole + early Hawking radiation + late Hawking radiation'' is

\bea \label{bel}
\psi_{B_2 EL} && = \alpha \left( \ \gamma \left| \Uparrow_{B_2} \ \downarrow_L \ \downarrow_E \right\rangle  + \delta \left| 0_{B_2} \ \uparrow_L \ \downarrow_E \right\rangle \ \right)   +  \nonumber \\
&& \beta \left( \ \rho \left| \Downarrow_{B_2} \ \uparrow_L \ \uparrow_E \right\rangle + \sigma \left| 0_{B_2}  \ \downarrow_L \ \uparrow_E \right\rangle \ \right)
\eea
The notation is self-explanatory. If a black hole was in the state $\uparrow_{B_1}$, and emits a spin down electron, it will go into the state with two units up  $\Uparrow_{B_2}$. The subscript on the black hole state, n on  $B_n$ represents how many decay modes have left the original black hole $B_0$. If a black hole was in the state $\uparrow_{B_1}$, and it emits a spin up electron it will go into the state with zero spin  $0_{B_2}$. Similarly, if a black hole was in the state $\downarrow_{B_1}$, and emits a spin up electron, it will go into the state with two units down  $\Downarrow_{B_2}$.

In fact we can put some restrictions on the values of the coefficients from the known facts about the black hole evaporation.  Detailed studies of the Hawking evaporation strongly suggest that black holes try very hard to get rid of their quantum numbers. Positively charged black holes preferably emit positively charged particles, while the emission of negatively charged particles is strongly suppressed.  Similarly, black holes with spin strongly prefer to emit particles with spin parallel to the spin of the black hole. This effect, which was first observed numerically in \cite{Page:1976ki} (see also \cite{Dai:2007ki}), was explained analytically in \cite{Dai:2010xp}.  Namely, what an observer at infinity observes depends on several factors, where only one is truly random.  The only random effect near horizon is vacuum fluctuation, where pairs of virtual particles are popping in and out of vacuum. The next step, when it is decided which of these virtual particles are going to become real on account of black hole's energy, and which of those created particles will end up crossing the potential barrier, are not random.  On the contrary, the black hole radiation is not random and strongly depends on the geometry and charge of the black hole. This is a very important fact to  notice, because by ignoring it one might derive that the mutual infirmation between the black hole state $B$ and the mode $L$ is zero, i.e. they are not entangled. This would then leave only the entanglement between $L$ and $E$, as assumed in the original firewall setup. However, we see that this can not be correct since it would contradict detailed numerical studies of Hawking evaporation from a black hole.

The Hawking radiation spectrum described by the energy $dE$ emitted by a rotating black hole per unit time $dt$ and per unit frequency $d\omega$ is
\be \label{shr}
 \frac{d^{2}E}{dtd\omega
}=\sum_{l,m}\frac{\omega }{e^{(\omega -m\Omega)/T_{h}}-(-1)^{2s}} \frac{N_{l,m}|A_{l,m}|^{2}}{2\pi} \, ,
\ee
where $T_{h}$ is the Hawking temperature, $l$ and $m$ are the total angular momentum quantum numbers, $s$ is the spin of the particle, $\Omega$ is the angular velocity of the black hole,  $N_{l,m}$ is the number of available degrees of freedom and $A_{l,m}$ is the absorption coefficient. The absorption coefficient $A_{l,m}$ actually determines the transmission cross section of a particle interacting with the black hole potential barrier, i.e. the probability that a created particle will penetrate the barrier. We can schematically represent Eq.~(\ref{shr}) as a product of two terms
\be \label{bbgb}
 \frac{d^{2}E}{dtd\omega
}=BB \times GB \, ,
\ee
where $BB$ stands for the thermal black body term, while $GB$ stands for the greybody term. The black body term gives the probability that a certain particle is produced near horizon as a real particle, while the greybody term modifies the thermal radiation due to the existence of the potential barrier which the created particle has to penetrate. The absorption coefficient $A_{l,m}$ crucially depends on the potential barrier and directly gives the greybody factor.  The black body term in Eq.~(\ref{bbgb}) preferably creates a particle of the same spin as the original black hole and the greybody factor shows that such particle will more likely cross the angular momentum barrier. Thus, the black body and greybody factors work in synergy.

A black hole with spin zero would prefer to emit particles with no spin, but in our example we have only electrons at our disposal. Such a black hole would not have any preference to emit either spin down or spin up electron. Therefore, the coefficients in the first step $\alpha$ and $\beta$ are of the same order. However, in the second step, the coefficients $\gamma$ and $\rho$ would tend to be very small since such processes are strongly suppressed. Thus we learn that
\be
\alpha \approx \beta , \ \ \   \gamma \ll \delta , \ \ \  \rho \ll \sigma  .
\ee

Our knowledge about black hole evaporation indicates that what an observer registers at infinity is not random at all. It is actually strongly correlated with the black hole itself. Thus, when we consider entanglement in the presence of the black hole, unitarity implies that we cannot neglect the state of the black hole.

To see more generally what unitarity of black hole evaporation implies, we recall that entanglement allows us to learn about one part of the entangled state by doing measurement on the other part. In the context of the manifestly unitary black hole decay process in Fig.~\ref{process} this implies the following. If we measure the state of $E$, and for example find it in the state $\left| \uparrow_{E} \right\rangle $, then we automatically know the state of the system $B_2 +L$, i.e.

\be \label{}
\psi_{B_2 L} = \lambda \left| \Downarrow_{B_2} \ \uparrow_L \right\rangle  + \delta \left| 0_{B_2} \ \downarrow_L \right\rangle .
\ee
Similarly, if we measure the state of $L$, and for example find it in the state $\left| \downarrow_{L} \right\rangle $, then we automatically know the state of the system $B_2 +E$, i.e.
\be \label{}
\psi_{B_2 E} = \alpha \left| \Uparrow_{B_2} \ \downarrow_E \right\rangle  + \beta \left| 0_{B_2} \ \uparrow_E \right\rangle .
\ee
Similarly, if we know the state of the black hole $B_2$, then we automatically know the state of the system $E+L$. This means that either $E$ is maximally entangled with $B_2 +L$; or $L$ is maximally entangled with $B_2 +E$; or $B_2$ is maximally entangled with $E +L$; but it never means that $L$ is maximally entangled with $E$ (or $B_2$)  separately as the firewall setup assumes.

\section{Icezone}
\label{I}

In the previous section we treated black hole evaporation as a quantum mechanical decay process and found that before a black hole has finished evaporating, if we single out the late mode $L$, all that we can say is that $L$ is entangled with a combination of early radiation $E$ and the black hole $B$.  Compared to a quantum decay process, the AMPS argument assumes that the late radiation $L$ is highly entangled with the early radiation $E$, and also that the late radiation $L$ is entangled with some interior mode separately.  In this section we apply Page's result at different times during the black hole evaporation process which supports our argument of the previous section.  As a result we argue that black holes form an icezone near the event horizon in the AMPS setup without invoking fundamentally high energy or non-local physics.

The firewall argument by AMPS used Page's result about the average entropy of a subsystem.  When a pure system is split into a larger subsystem, $E$, and a small subsystem, $L$, then typically the smaller subsystem $L$ is maximally entangled with $E$.  This result has been used to argue that if one waits long enough, a time called the Page time, then most of the degrees of freedom are in the already radiated early Hawking radiation, $E$.  The subsystem $E$ is then maximally entangled with the rest of the system, which they claim to be the late radiation $L$.   We will argue that Page's result when applied to a small subsystem $L$ should also include early radiation $E$ and the late radiation that we call $L_R$. We can ignore the black hole only at the end of evaporation when the black hole does not exist, but at any earlier time unitarity arguments that we explored in the previous section imply that the black hole can not be ignored.

AMPS also claimed that the smoothness of the horizon implies that $L$ is entangled with some mode $L^\prime$ inside the black hole.  A physical reason would be that Hawking radiation arises from a pair of virtual particles being created near the event horizon.  The outgoing mode and ingoing mode would be entangled with each other due to their common origin.  Another reason to expect such an entanglement is that the event horizon of a black hole is classically approximated by Rindler space.  Quantizing Rindler space entangles the modes inside and outside the horizon.   In this section though we argue that quantizing Rindler space is not a good approximation for old black holes, and the fact that $L$ is entangled with a combination of early radiation $E$ and the black hole $B$ implies that some kind of an extended entanglement is needed which goes beyond the local Rindler approximation.

Let us now apply Page's result and start by examining things at the end of evaporation when there are no black holes or remnants around and we have to make the fewest assumptions.  At this moment the radiation, $E$, emitted before the Page time and the rest of the radiation, $L_R$, emitted after the Page time are maximally entangled.  It is important to note that $L_R$ is not just a single mode of radiation.  Evolve the late radiation $L_R$ back in time to when the black hole existed.  This radiation is not only blueshifted to some radiation wavepacket, $L_0$, but some of the radiation was organized into black hole degrees of freedom $B$.  The black hole state and radiation wavepacket $B+L_0$ evolves into the late $L_R$.  At this point one could argue that the wavepacket $L_0$ is maximally entangled with other degrees of freedom, $L^\prime_0$, but we see there is no paradox about non-monogamous entanglement.  The late radiation $L_R$ is maximally entangled with E at late times when the black hole has evaporated, while $L_0$ is entangled with $L'_0$ when the black hole is still present.  The difference between this discussion and the AMPS discussions can be traced back to the inclusion of backreaction.  Strictly speaking the propagation of radiation in the black hole geometry can not be described by free propagation in a fixed background.  The late radiation propagates in a spacetime but when we backwards time evolve this radiation, there is a black hole present and this backreaction affects the entanglements of the states.  Reference \cite{Verlinde:2013uja} also points out the difference between a virtual mode near horizon and a real mode observed at infinity, though in slightly different setup.

\begin{figure}
   \centering
\includegraphics[width=10cm]{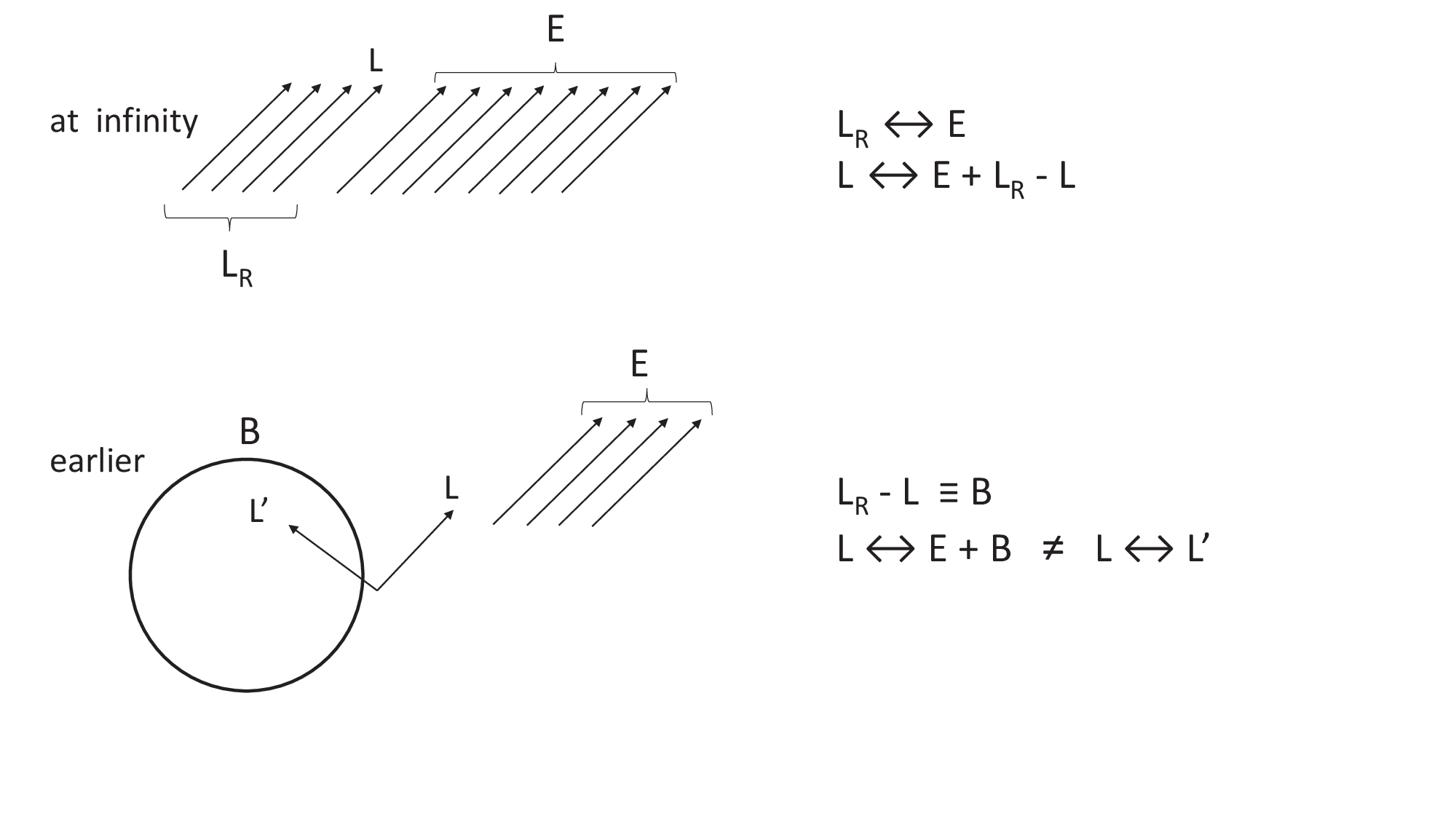}
\caption{At the end of evaporation, the late radiation $L_R$ is maximally entangled with the early radiation $E$.  Singling out a mode $L$ of the late radiation, Page's result implies that $L$ is highly entangled with the rest of the radiation $E+L_R-L$.  Time evolving the radiation back in time to when $L$ was just emitted, the system consists of the mode $L$, early radiation $E$ that was already emitted, and an old black hole.  At early times, the late radiation was organized as a black hole, i.e. $L_R-L\equiv B$.  The mode of late radiation, $L$ is entangled with the early radiation and black hole, $L \leftrightarrow E+B$, instead of with just a mode $L^\prime$ inside the horizon .  Due to the large macroscopic entanglements of $L$, Rindler space can not be good semiclassical description of the event horizon for old black holes.}
\label{LE}
\end{figure}

One can also focus on the single wavepacket, $L$, of the late radiation $L_R$.  At the end of evaporation $L$ is maximally entangled with the rest of the radiation, $L\leftrightarrow (E+L_R-L)$, where the two-sided arrow means maximally entangled as illustrated in Fig.~\ref{LE}.
Time evolve the radiation backwards to near the Page time when there is an old black hole around but the mode $L$ and early radiation $E$ are still outside the horizon.  The late radiation excluding one mode, $L_R-L$,  was previously organized into the degrees of freedom of the black hole.  All together we find that the mode $L$ is entangled with the early Hawking radiation and the black hole by Page's result, $L\leftrightarrow E+B$.  This result is different from the one claimed in the firewall paradox $L\leftrightarrow L^\prime$.  Even though $L_R\leftrightarrow E$, it does not have to be true that $L$ a subset of $L_R$ is entangled with a subset of $E$.

For old black holes, the entanglement of $L$ with a combination of both the early Hawking radiation and black hole states suggests that the quantum description of a black hole changes over time.  The quantum Rindler approximation of a black hole horizon can not be valid because that would require the inside and outside modes to be entangled with each other.  Our result is similar to having a firewall in the sense the black hole horizon is no longer a naive model of flat space but we argue that entanglements are modified in a very different way.  To recap our claim is:

\vspace{.1in}
{\centering{\it{The quantum Rindler approximation breaks down near the horizon of an old black hole due to unitary evolution.}}}
\vspace{.05in}

\noindent  If one started with the assumption that a quantum Rindler horizon is an accurate description of an old black hole then such a macroscopically large entanglement would appear to violate locality.  As we will argue below, however, there appears to be a local dynamical mechanism that enforces this entanglement.

Let us continue the backward time evolution to when we have a young black hole and only some portion of the early radiation $E$ radiation has been released.  The portion of the radiation, $E^\prime$ that is outside the horizon, is a small subsystem.  By Page's result $E^\prime$ should be maximally entangled with the rest of the larger system which is the black hole.  So for young black holes, the modes outside the horizon $E^\prime$ are strongly entangled with the modes of the black hole inside the horizon and the quantum Rindler approximation should be accurate.

\begin{figure}
   \centering
\includegraphics[width=9.1cm]{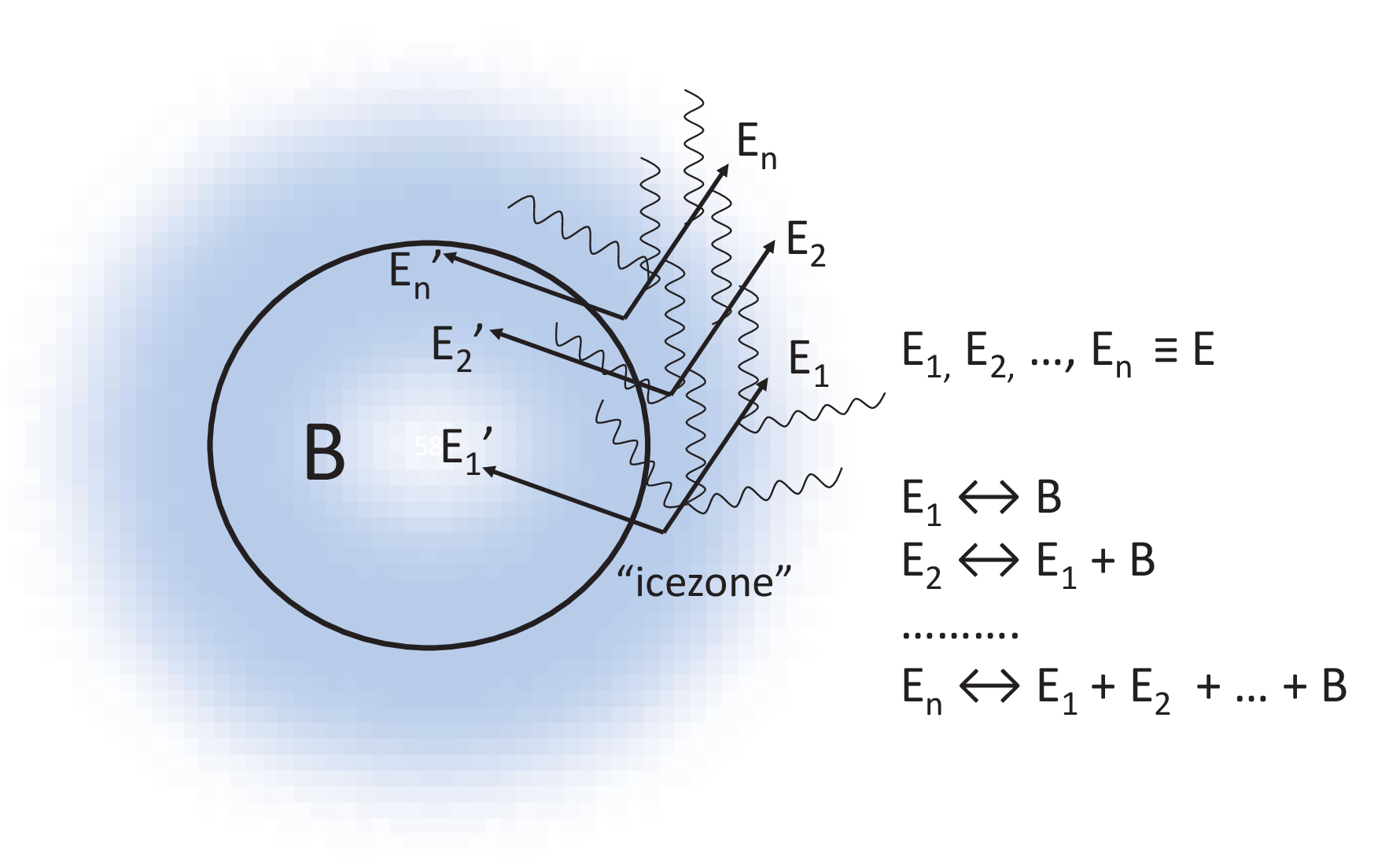}
\caption{$E_1, E_2, ..., E_N$ are the modes of early Hawking radiation.  A local Rindler approximation would naively imply that each $E_i$ is entangled  with its ingoing partner $E'_i$. However unitarity implies  that the first mode $E_1$ is entangled with the black hole, B, the second mode $E_2$ is entangled with the system of $E_1+B$, etc. All the outgoing modes will eventually be entangled. The mechanism that provides the necessary interaction between the modes to establish the entanglement is the ``icezone'', which consists of a sea of quanta that mediate interactions (either perturbative or non-perturbative) between the thermal Hawking quanta.}
\label{icezone}
\end{figure}

Although the above general arguments suggest a modification to our usual black hole understanding, there should also be a local dynamical process that realizes how the entanglement changes at large scales from the horizon.  Our proposal is to modify the usual physical explanation of Hawking radiation in which two virtual particles pop up from the vacuum with one ingoing and one outgoing mode.  It is much more accurate to say that the two virtual particles will appear along with a large number of soft infrared photons (or other low energy particles).  These low energy particles will collide with Hawking quanta in the later radiation as shown in Fig.~\ref{icezone}. We emphasis that any black body emitters, and in particular black holes, emit a large number of infrared modes which can be seen from the distribution function for the number of emitted particles per unit time per unit frequency band $dN/(dt d\omega)=1/(e^{\omega/T}-1)$, where $\omega$ is the frequency of the emitted particles and $T$ is temperature of the emitter. Obviously, when $\omega \rightarrow 0$, the number of emitted modes, $N$, diverges (see also Eq.~(4) in \cite{Page:1976df}). The point is that the flux that these infrared modes carry is negligible (to get the energy flux we have to multiply the distribution with extra powers of $\omega$), so it does not change the Hawking’s result for the emitted flux at infinity. Nevertheless, these modes mediate interactions which can redistribute the entanglement.
Due to the collisions of the main Hawking quanta with these infrared modes, which may also include loop diagrams with different particles of the theory, a late mode $L$ is not only entangled with the interior modes of the black hole but with modes that are in the Hawking radiation E.  Low energy particles would not measurably affect the amount of radiation being emitted (either because they are very soft or they are suppressed by loops) but they would change the entanglement.  Due to the interaction in the radiation, the entanglement of the horizon modes will spread out spatially to the far away radiation.

Although the standard Hawking radiation is usually defined for a static observer far from a black hole, it is important to emphasize that an infalling observer who probes the near horizon structure would certainly notice these soft quanta.  For example it was explicitly demonstrated in \cite{Greenwood:2008zg} that the infalling observer sees a non-divergent flux of particles coming out of the black hole. Though an infalling observer should classically encounter a nearly flat space-time at the horizon of a large black hole, these soft quanta would affect entanglement and ruin the quantum Rindler approximation.

There is so far no apparent reason to argue that a firewall is needed to solve the discrepancy between the usual black hole evaporation process and quantum mechanics. The real problem was the formulation of the paradox in \cite{Almheiri:2012rt}, which considers the entanglement of the black hole states with late Hawking radiation, and late Hawking radiation with early Hawking radiation separately.  This formulation is not self-consistent because the a mode of late Hawking radiation is not entangled with either the black hole or early Hawking radiation separately. Instead, the unitarity implies that it is entangled with the combined ``black hole and early Hawking radiation''. This extended entanglement can be easily explained with small corrections to the thermal radiation that come from interactions which are neglected in the original Hawking derivation.

\subsection{Quantifying the icezone effect: density matrix}
\label{dm}

As an example of how icezone works in practice we study a particular spin emission model of black hole evaporation where interactions change an approximately thermal density matrix into a pure state.  These interactions introduce small corrections to Hawking radiation during most of the decay process but become important at the end of evaporation and unitarize the radiation.


We model our black hole as a massive object emitting spin $1/2$ quanta.  Initially the number of quantum gravitational quanta contained in a black hole is of the same order as its entropy, i.e. $N=(M/M_{Pl})^2$, where $M$ is the mass of the black hole (this number $N$ does not include the soft interaction modes between them since they do not significantly change the flux at infinity). For each emission, emitted quanta can be have one of the two spin values.  Therefore the space of states is spanned by $2^N$ basis vectors.  We analyze the density matrix of the radiation after $K$ quanta have been emitted.  The radiation is entangled with the states of the black hole and so while the black hole exists, they should not constitute a pure state.

Corrections to the density matrix due to interactions can be determined in the interaction picture as
\be
\rho(t)= G(t) \rho_0 = (\mathbb{I}+ A) \rho_0
\ee
where $\rho_0$ is the initial density matrix, while $G$ is a superoperator which is determined by the interaction Lagrangian..  For the black hole, the initial density matrix after $K$ quanta are emitted should be maximally entangled and close to Hawking's result
\be
\rho_0= \frac{1}{2^K} \mathbb{I}_K
\ee
where the factor of $1/2^K$ is a normalization factor to ensure that the probability is normalized.  The general rule of thumb is that such a density matrix needs another factor of $1/2$ for each additional quanta emitted.  When a virtual pair is created near the horizon, one member of the pair sinks into the black hole and the other
leaves to infinity. One traces over the state which is inaccessible and the density matrix  is $2$-dimensional unit matrix with the normalization $1/2$ to preserve the sum of
probabilities. When the next pair is emitted, the matrix is $4$-dimensional and normalization is $1/4$, so after $K$ steps, the normalization is $1/2^K$.

In the second equality we used that fact that because $G$ is a time evolution operator, it should be equal to the identity matrix plus some interaction matrix $A$.  In general the interaction matrix should depend on time but we are interested in analyzing the situation after a sufficiently long time has passed and after a large number of interactions.  The interaction matrix connects different states, so it is off-diagonal.  For simplicity we set the many possible phases equal to zero.  We also assume that interactions are random and equally mix all possible states.  This constrains the form of the interaction matrix so it is one everywhere except for zero along along the diagonal
\be
A=a \cancel{\mathbb{I}}=a
\begin{bmatrix}
0&1&1 &... \\
1&0&1 &... \\
1&1&0 &...\\
...&....&...&...
\end{bmatrix} \ .
\ee
The form of the coefficient $a$ can be determined from perturbative or non-perturbative interactions depending on the theory in question.  If the effect is non-perturbative, then it is expected that the magnitude of the corrections scales as
\be \label{npc}
a=e^{-S} \approx 1/2^{N-K} .
\ee
This is the probability that non-perturbative effects take place, and this probability obviously increases as more quanta are released.  After $K$ quanta have been radiated
\bea
\rho_{K}&=& (\mathbb{I}_{K}+ \frac{1}{2^{N-K}} \cancel{\mathbb{I}}_K) \frac{1}{2^K} \mathbb{I}_K\\
&=&\frac{1}{2^K}\mathbb{I}_K +\frac{1}{2^N}\cancel{\mathbb{I}}_K \\
&=&\rho_0+\frac{1}{2^N}\cancel{\mathbb{I}}_K \ .
\eea
In this expression we see that there are two sources that are suppressing the effect of interactions.  The first is the normalization of the density matrix which effectively means that effect of interactions is effectively  suppressed by the number of possible final states.  This suppression grows as the black hole evaporates.  The second suppression is due to the probability of each non-perturbative interaction.  This suppression starts off as very large but becomes much smaller as the black hole evaporates.  For large  $N$, the interactions are initially suppressed compared to the Hawking radiation due to the non-perturbative interaction.  As the black hole radiates away though, the suppression factors become equal and we get
\be \label{dmnp}
\rho_N=  \frac{1}{2^{N}} (\mathbb{I}_{N}+\cancel{\mathbb{I}}_N) = \rho_{\rm Pure}
\ee
which is just the density matrix $\rho_{\rm Pure}=\frac{1}{2^{N}} \sum_{i,j} |i><j| $ of a pure state $\psi=\frac{1}{2^{N/2}} \sum_i |i>$.

This final result that non-perturbative corrections to Hawking radiation can unitarize Hawking radiation is similar to that of \cite{Papadodimas:2012aq}.  Even though the starting points are very different and our follows by analyzing the interaction picture of time evolution, one can show that the diagonal correction matrix in \cite{Papadodimas:2012aq} is equivalent (by a change of basis) to our off-diagonal matrix
\be
\begin{bmatrix}
2^K-1&0&0&... \\
0&-1&0&... \\
0&0&-1&...\\
...&...&...&...
\end{bmatrix}
=M
\begin{bmatrix}
0&1&1&... \\
1&0&1&... \\
1&1&0&...\\
...&...&...&...
\end{bmatrix} M^{-1}\ .
\ee

Non-perturbative gravity effects in principle could solve the black hole information problem.  Their effect could be very important in certain theories especially if the radiated quanta are from a free field.  For different theories, however, perturbative effects could be more dominant.  In those cases the interaction matrix should take the form
\be  \label{im}
A=\frac{1}{\frac{N!}{(N-K)! K!} }\cancel{\mathbb{I}}
\ee
which is normalized by the number of ways to choose $K$ quanta out of $N$ total quanta.
The reason for this form in that the interaction matrix can be viewed as a multi-particle state which has to be properly normalized. A black hole of a given mass can emit a certain number $N$ of particles during its lifetime.  At any given finite moment, after $K$ quanta are emitted, the matrix contains only $K$ terms out of possible $N$, so the multi-particle state should be normalized by the number of ways to choose $K$ quanta out of $N$ total quanta.

In this case we can check several limits for the density matrix
\be
\rho_K=(\mathbb{I}_{K}+\frac{1}{\frac{N!}{(N-K)! K!} }\cancel{\mathbb{I}}_K) \frac{1}{2^K} \mathbb{I}_K = \rho_{0} +  \frac{1}{2^K \frac{N!}{(N-K)! K!}} \cancel{\mathbb{I}}_K \ .
\ee
At the end of the evaporation process $K=N$ we get $A=\cancel{\mathbb{I}}_N$. This tells us that at the end of the decay process we obtain
\be \label{dmp}
\rho_N=\frac{1}{2^N} (\mathbb{I}_{N}+\cancel{\mathbb{I}}_N) \equiv \rho_{\rm Pure}
\ee
which is the pure state density matrix just like before.  For $K=1$ we get $A= \frac{1}{N}\cancel{\mathbb{I}}$ so the corrections are suppressed by the large $N$ factor.  With this form of the corrections, the suppression grows until we get to $K=N/2$ and then the suppression decreases.  This result is in fact closer to Page's result which states that the black hole only starts to emit information at significant rate after half of the black hole has been radiated away.

The above discussion agrees with the qualitative features of the icezone discussed in the previous section. The icezone takes time to form.  When only a small number of Hawking quanta has been released then the effect of the interactions is small and the quantum Rindler space is a good approximation of the horizon.  After a long time, many interactions between the emitted quanta which can either perturbative or non-perturbative depending on the theory, will noticeable change the entanglements near the horizon.  The quanta mediating interaction between the thermal Hawking quanta  would not change the flux of the outgoing quanta or, the energy enough to significantly change the classical nature of the horizon.  These interaction would change the fine quantum structure of the horizon.  The dynamics of interactions  would not be a firewall but the exact opposite, an icezone. The main point is that each individual interaction term makes tiny corrections to the Hawking radiation but over time their integrated effect is large (see also \cite{Saini:2015dea}).

\section{A simple model}

It was important to learn that both non-perturbative and perturbative interactions could purify the thermal density matrix. Otherwise, the resolution to the information loss paradox would strongly depend on the details of the underlying theory. In principle, the resolution should not depend on the exact particle content of the standard model and its interactions.  The resolution should work even for a simple generic model. Say for example that we have only a single massless scalar field, $\phi$, propagating in the  background of a black hole. Its Lagrangian density is
\be
L = \frac{1}{2} g^{\mu \nu} \partial_{\mu} \phi  \partial_{\nu} \phi \ .
\ee
However, gravity would induce many other terms even if they were initially absent from the original bare Lagrangian.  The gravitational induced terms would modify the action so it is of the form
\bea \label{np}
L &=& \frac{1}{2} g^{\mu \nu} \partial_{\mu} \phi  \partial_{\nu} \phi + M_{\rm Pl}^3 \phi +M_{\rm Pl}^2 \phi^2 + M_{\rm Pl} \phi^3 + \frac{1}{M_{\rm Pl}^0} \phi^4  \nonumber \\
& &+  \frac{1}{M_{\rm Pl}} \phi^5\, + \ldots
\eea
Terms of the form $\phi^n$ for $n\geq 5$ are called perturbative.  Perturbative terms are suppressed by powers of $M_{\rm Pl}$, so they are small (except for the dimension four term). Non-perturbative terms are of dimensions one, two and three.  These terms are not suppressed by powers of $M_{\rm Pl}$, but in addition to the factors of $M_{\rm Pl}$ written explicitly above, they also involve non-perturbative quantum gravity effects (topology change, wormholes, etc.) and are suppressed by large action factors $e^{-S}$, as in our previous discussion (see also \cite{Stojkovic:2013ppa,Kallosh:1995hi}). Thus, even a simple model of the massless scalar field has enough structure in the presence of gravity to yield both non-perturbative and perturbative interactions.

We can now apply general arguments from the sections \ref{I} and \ref{dm} to the simple model in Eq.~(\ref{np}). Consider a single scalar field like in Eq.~(\ref{np}) propagating in the gravitational background of a black hole, i.e. $g^{\mu \nu}$ is that of a black hole. A black hole of mass $M$ will excite $N=(M/M_{Pl})^2$ leading order Hawking quanta.
In addition, we have modes which will mediate interactions between them.
If the interactions between the leading order Hawking quanta are governed by $\phi^2$ and $\phi^3$ terms in Eq.~(\ref{np}), then they will be non-perturbative. For example, the term  $\phi^2$ directly couples two Hawking quanta, while the term $\phi^3$ directly couples three Hawking quanta. Since these are non-perturbative effects, the magnitude of the corrections to the diagonal Hawking's density matrix scales as $a=e^{-S}$, where $S$ is the entropy of the black hole. For a large black hole, these interactions are initially strongly suppressed, but they grow as the black hole evaporates and its entropy goes down. At the end of evaporation, the resulting density matrix will be that of a pure state, as we can see from Eq.~(\ref{dmnp}).

If the interactions between the main Hawking quanta are governed by the terms in Eq.~(\ref{np}) whose dimension is four and higher, then they will be perturbative interactions.
In that case the magnitude of the corrections scales as $a= \left(\frac{N!}{(N-K)! K!} \right)^{-1}$, which is just a combinatorics factor to choose $K$ quanta (emitted at a given moment) out of $N$ total Hawking quanta that a black hole can emit during its lifetime. This is the proper normalization of the interaction matrix in (\ref{im}) which can be viewed as a multi-particle state. In principle, the higher  perturbative terms have further suppression from powers of $M_{\rm Pl}$, but the $\phi^4$ term is free of this suppression.
Again, for a large black hole, the contribution of these perturbative interactions to the total density matrix is initially strongly suppressed, but it grows as the black hole evaporates and $N$ approaches $K$. At the end of evaporation, when $N=K$, all the emitted Hawking quanta are strongly entangled by interactions, and the resulting density matrix will be that of a pure state, as we can see from Eq.~(\ref{dmp}).

\section{Conclusions}
In this paper we revisited some of the assumptions in the formulation of the firewall paradox. The heart of the paradox is the apparent expectation that a mode of late radiation near the black hole horizon should be maximally entangled with both a mode inside the horizon and the early radiation mode, which is impossible in quantum mechanics.  Instead of trying to destroy entanglements to prevent a paradox we argued that it would be more conservative to allow for a gradual change in the entanglements due to the accumulated effect of low energy physics.

In Sections \ref{u} and \ref{I}, we pointed out that the claim of early and late Hawking radiation being maximally entangled due to unitarity is not precise enough.  When an old black hole is present, the early radiation must be maximally entangled with the system of the late mode plus the black hole, and not separately with each of them, as we illustrated in Fig.~\ref{process}.  One of the reasons why the authors of the firewalls paradox derived that the late mode $L$ is maximally entangled with the early radiation $E$ is that they assumed that the black hole radiation is random, which enforces the mutual information between the black hole state $B$ and the late mode $L$ to be zero. However, this assumption is not correct and contradicts detailed numerical studies of Hawking evaporation from a black hole. Namely, what an observer at infinity observes depends on several factors, where only one is truly random.  The only random effect near horizon is vacuum fluctuation, where pairs of virtual particles are popping in and out of vacuum. The next step, when it is decided which of these virtual particles are going to become real on account of black hole's energy, and which of those created particles will end up crossing the potential barrier, are not random.  This, the black hole radiation is not random and strongly depends on the geometry and charge of the black hole. Thus, the states $B$, $L$ and $E$ are mutually all entangled.

Due to the extended nature of the entanglements, the horizon is not always well described by a quantum Rindler space.  In addition, the late mode at the end of evaporation is not the same as the late mode at the moment when it was emitted.  As a black hole ages, its horizon is slowly modified in a quantum mechanical sense due to interacting low energy modes forming an icezone as shown in Fig.~\ref{icezone}.  Using only local physics this icezone increases the entanglement over larger and larger spatial distances as the black hole evaporates and it would be interesting to explore connections with the wormhole picture described in Ref.~\cite{Maldacena:2013xja}.

The ``icezone" interactions are not strong (they are mediated by infrared modes which are present in large numbers in any black body emitter), so they don't change the acceleration or trajectory of the outgoing mode too much.  For large black holes this is a relatively slow process because the radiation flux is so low, but after a long time the accumulation of infrared collisions substantially changes the entanglement near the horizon.

To quantify the icezone effect we analyzed a spin emission model of quanta from black holes in Section \ref{dm}.  We explicitly constructed the density matrix for the icezone for non-perturbative and perturbative effects.  The construction was based on the interaction picture of interactions and in the non-perturbative case we found similar results to \cite{Papadodimas:2012aq}.   The main point is that while these interactions are typically small, they become large and of the order of the Hawking radiation when the black hole has almost finished evaporating.  The effect eventually purifies the thermal radiation and information is recovered from the black hole. This does not imply that independently of the initial state of the full system, the final state is pure.
In particular, a mixed initial state will not evolve into a pure state. Interactions in the background geometry
of the black hole can un-mix only what this black hole has apparently mixed.  As we extensively discussed in Section \ref{u}, processes near the black hole strongly
depend on the black hole geometry, and one cannot logically expect that mixing that occurred in the preparation of the initial state prior to the black hole formation
(say by tracing out some of the degrees of freedom of the initial state) will be unmixed in the background of the black hole, simply because these two processes are unrelated.

We emphasize that an icezone is observable even for an infalling observer who is able to probe the near horizon structure. This is important since the standard Hawking radiation is usually defined for a static observer far from a black hole. For this far away observer, the blueshifted flux of radiation diverges at the horizon. This is also the reason why a static observer near the horizon sees entanglement degradation.  However, as explicitly demonstrated in \cite{Greenwood:2008zg} (see also \cite{Barbado:2012pt,Kim:2009ha}), an infalling observer would also observe a flux of particles coming out of the black hole, with an important difference being that this flux would be non-divergent at the horizon. Thus, an infalling observer should indeed classically encounter a nearly flat space-time at the horizon of a large black hole, however these soft quanta would affect entanglement as we described it and invalidate the quantum Rindler approximation.

Icezone, which is a quantum effect in vicinity of a black hole, changes how the black hole states are entangled by modifying the quantum near horizon limit.  It would be interesting if it could be observed experimentally.  For a real black hole accretion disk, the effects of icezone would typically be small due to its low temperature.  Ideally, there may be very special cases (e.g. an isolated black hole) in which icezone could be experimentally observed by interference measurements.

Finally, we point out the crucial difference between the firewalls and icezones. Unlike firewalls, icezones retain the rules of QM and thermodynamics and change general relativity to conform to them instead of the other way around.  We then see that there is no paradox from the beginning or need to introduce unexplained high energy radiation appearing out of nowhere.

For a partial list of related work see \cite{Almheiri:2013wka,Kim:2013caa,Dundar:2014gma,Mann:2014yxa,Brown:2015yma,Dundar:2015axd,Kim:2016iyf,Mann:2015luq,Ho:2015vga,Dai:2016sls,Dai:2016jdv,Saravani:2012is}.

\begin{acknowledgments}
The authors are thankful to De Chang Dai, Raphael Bousso, Don Marolf, Don Page and Herman Verlinde for very useful discussions. This work was partially supported by NSF, grant numbers PHY-1066278 and PHY-1417317.
\end{acknowledgments}

\end{document}